\input{psfig}                                  

\def\ltae{\raisebox{-0.6ex}{$\,\stackrel
{\raisebox{-.2ex}{$\textstyle <$}}{\sim}\,$}}

\def\Oiii{{[O{\,\small III}]~$\lambda$5007}\/}
\def\Oii{[O{\,\small II}]~$\lambda$3727\/}

\def\O4363{[O{\,\small III}]~$\lambda$4363\/}

\def\Ho{{$H_0$}\/}

\def\kmsMpc{{km~s$^{-1}$~Mpc$^{-1}$}\/}
\def\ergs.cm2.s{{ergs~cm$^{-2}$~s$^{-1}$}\/} 

\def\arcdeg{{^{\circ}\!\!}\/} 

\def\etal{{ et~al.\ }}

\def\smallskip{\vskip 6pt}

\documentclass{aa}
 
\begin{document}
\thesaurus{ 06() }  

\title{\Large\bf The radio structures of southern 2-Jy radio sources: \\ 
new ATCA and VLA radio images.}

\author{R. Morganti\inst{1,2}\thanks{morganti@nfra.nl}, T. 
Oosterloo\inst{2}, C.N.  Tadhunter\inst{3}, R.  Aiudi\inst{1}, P. 
Jones\inst{4}, M. Villar-Martin\inst{5}}

\institute{Istituto di Radioastronomia, CNR, Via Gobetti 101, 40129 Bologna,
Italy \and Netherlands Foundation for Research in Astronomy, Postbus 2, 7990
AA Dwingeloo, The Netherlands \and Dept.  Physics, University of Sheffield,
Sheffield S3 7RH, UK \and Centre for Astronomy, University of Western Sydney,
Nepean, PO Box 10, Kingswood, NSW 2747, Australia \and Institute d'
Astrophysique de Paris (IAP), 98 bis Bd Arago, F75014 Paris, France}

\offprints{R.Morganti}

\maketitle

\markboth{The radio structures of southern 2-Jy radio sources}{Morganti et al.}

\begin{abstract}

We present new radio images obtained with the Australia Telescope Compact
Array (ATCA) and the Very Large Array (VLA) for a group of 14 galaxies
belonging to the 2-Jy sample of radio sources.  The new images improve the
data already available on these objects and, in general, the database that we
are building up on the sample.  They will also be used for follow-up work
where radio-optical comparison will be done.  \\ We briefly discuss the core
dominance parameter ($R$) for the objects for which the new data have given
new information and, in particular, for broad line radio galaxies (BLRG).  
One of the BLRG does not show a core, even at 3~cm, and this is at variance
with the general tendency of BLRG to have relatively strong cores. 
The depolarization is also discussed for a group of small double-lobed radio
galaxies. 

\end{abstract}


\section {Introduction}

Multiwaveband studies of Active Galactic Nuclei (AGN) have become an essential
part in understanding these objects.  Indeed, our knowledge  of this group
of objects is not complete until the relation between their (e.g.) optical,
radio and X-ray emission is clear.  Moreover, part of the study of AGN
concentrates on the importance of `unified schemes' in explaining the
appearance of these objects.  Tests for these schemes are more efficient when
different characteristics in different wavebands have to be explained. 

For these reasons we started a few years ago the study of a complete
sample of radio sources with the idea of building an homogeneous
database that could allow a comprehensive investigation of various
aspects and open questions connected with AGNs.  For this study we have
used the southern part of the so-called 2-Jy sample (Wall \& Peacock
1985) with selection criteria $S_{2.7GHz}> 2$ Jy, $\delta<+10^\circ$ and
$z<0.7$.  This sample includes 88 radio sources (radio galaxies and
quasars) of many different classes, i.e.  Fanaroff-Riley (Fanaroff \&
Riley 1974) type I and II, core/halo, Compact Steep Spectrum (CSS) and
flat spectrum sources.  For all these objects we have now collected
optical spectra and polarimetry (Tadhunter \etal\ 1997, Tadhunter et al. 
1998), X-ray data (Siebert \etal\ 1996, Padovani et al.  1999, Trussoni
et al.  1999) and radio data.  The latter comprise images (VLA and ATCA,
Morganti \etal\ 1993) and high resolution core flux density measurements
(Morganti \etal\ 1997a). 

A number of interesting results have been already obtained from this
database (see Morganti \& Tadhunter 1997, Morganti et al.  1997a,
Tadhunter et al.  1997, Tadhunter et al.  1998 and ref.  therein). 
However, together with the results some new questions have arisen which
would benefit from higher quality data than available at
present.  In particular, for a number of sources the radio maps
available  are not good enough for a proper
morphological classification and to derive other important parameters.

In order to improve the available radio images we have carried out new
observations, for a number of radio galaxies in the sample, using either the
VLA or the ATCA, according to the declination of the sources.  This work can
be, therefore, considered as a follow-up and completion of the radio study of
this sample presented in Morganti et al.  1993.  In this paper we present the
new total intensity and polarisation radio images obtained for 14 radio
galaxies.    The objects selected are very
different in morphology: among them there are both small double-lobed and
large objects that needed a better radio image for a proper classification. 
These images will be mainly used for follow up work where
radio-optical comparison will be done.

Throughout this paper we assume that \Ho = 50 \kmsMpc and $q_o=0$.

\begin{table*}
\voffset=1truecm
\begin{center}
{\bf Table 1:  General characteristics}

\medskip
\def\mc{\multicolumn{1}{c}}
\def\mcc{\multicolumn{2}{c}}
\def\mccc{\multicolumn{3}{c}}

\begin{tabular}{lcccccc} \hline\hline
 Object   &      & $z$  & $\alpha^{4.8GHz}_{2.3GHz}$ &  $\log P_{tot}^{6cm}$ & 
LAS  & LLS \\
          &      &      &     & W Hz$^{-1}$  & arcsec
& kpc \\
\hline
         &       &       &       &        &       &     \\
0034--01 & 3C15   & 0.073 & 0.79  & 25.59  & 42 &  80  \\
0035--02 & 3C17   & 0.220 & 0.72  & 26.81  & 38 &  181 \\
0039--44 &        & 0.346 & 0.93  & 26.93  & 9.5 & 62  \\
0043--42 &        & 0.116 & 0.87  & 26.28  & 160 &  460 \\
0117--15 & 3C38   & 0.565 & 0.90  & 27.54  & 14  &  120 \\
0409--75 &        & 0.693 & 0.86  & 28.18  & 9   &  85  \\
0442--28 & OF-271 & 0.147 & 0.93  & 26.38  & 105 &  367 \\
0453--20 & OF-289 & 0.035 & 0.73  & 25.00  & 27  &  26  \\
0625--53 &        & 0.054 & 1.17  & 25.40  & 90  & 130 \\
1602+01 & 3C327.1 & 0.462 & 1.07  & 27.21  & 19  &  147 \\
1938--15 & OV-164  & 0.452 & 0.82 & 27.47  & 6.5  & 50  \\
1954--55 &        & 0.060 & 0.78  & 25.58  & 360  & 576  \\
2058--28 &        & 0.038 & 0.74  & 25.11  & 600 & 628   \\
2314+03 & 3C459  & 0.220 & 0.97  & 26.54  &   10  & 48   \\
         &       &       &       &        &      &       \\
\hline\hline
\end{tabular}
\end{center}  
\end{table*}

\section{Observations}

We have observed 14 radio galaxies with either the Australia Telescope Compact
Array (ATCA) \footnote{Operated by the CSIRO Australia Telescope National
Facility.} or the Very Large Array (VLA) \footnote{The National Radio
Astronomy Observatory is operated by Associated Universities, Inc., under
contract with National Science Foundation.} depending on their declination. 
One exception is 1938-15, which was observed with both the instruments. 
The general characteristics of the observed objects are listed in Table~1. 
The total power, calculated at 4.8~GHz, and the spectral index, estimated
between 2.7 and 4.8 GHz ($\alpha^{4.8GHz}_{2.3GHz}$, defined as $S \propto
\nu^{-\alpha}$), have been taken from Wall \& Peacock (1985). 

Large angular (LAS) and linear (LLS) sizes (taken from the lower resolution
images available in literature, see notes to the sources) and measured from the lower
contour, are also listed.

\subsection{ATCA observations}

Seven objects in the most southern part of this sample ($\delta < -30\arcdeg$,
with the exception of 1938-15) were observed with the ATCA.  These 
observations were done on the 4-7 April 96 in the 6km (6A) configuration and
on the 8 May 96 in the 1.5km (1.5D) configuration.  Data were taken
simultaneously at 4.8 GHz (6~cm) and 8.6 GHz (3~cm), each with a bandwidth of
128 MHz. 

The 3~cm observations (with the 6km configuration) are essential to improve
the previously available radio images and to be able to detect, e.g., jet
structures (if any) close to the core.  However, for the three larger sources
(0043--42, 0625--53 and 2058--28) we have also obtained observations with the
1.5 km configuration in order to better image the extended emission.  The data
from the two configurations have been combined to obtain the best images.  For
two objects (0043--42 and 2058--28) the resolution of our combined 3~cm data
was still too high to properly image the extended emission.  Only the 6~cm
data will be presented here for 2058--28, while for 0043--42 we only show an
image of the hot-spots at 3 cm.

Each source was observed in scans of about 20min for a total of about 6h in
each configuration.  The scans were spread throughout a 12 hour observing
period in order to optimise the $(u,v)$ coverage within the available
integration time.  The ATCA observing parameters are summarised in Table~2.

The ATCA data were calibrated by using the MIRIAD package (Sault, Teuben \&
Wright 1995), which is necessary for the polarisation calibration of ATCA
data.  The flux density scale was calibrated against observations of
PKS~1934$-$638, assumed to be 5.83 Jy at 4.8 GHz and 2.84~Jy at 8.6 GHz
according to the latest analysis by Reynolds (1996).  A few iterations of
phase-only self-calibration were then applied to each source.  Together with
the total intensity images we have also obtained the images for the Stokes
parameters ($Q, U$), the polarised intensity ($P=(Q^2+U^2)^{1/2}$) and
position-angle ($\chi= 0.5 {\arctan} (U/Q)$) images.  The polarised intensity
and the fractional polarisation ($m=P/I$) were estimated only for the pixels
for which $P>3\sigma_{QU}$.

In order to compare the 8-GHz and the 5-GHz data (being taken with the same
array they have different resolution), we have generated 8- and 5-GHz Stokes $I,
Q$ and $U$ images at the best 5-GHz resolution (see Table~2). This was done by
convolving the 8-GHz visibilities with the appropriate Gaussian during imaging,
thus weighting down the longer spacing.
The derived depolarization is defined as the DP$^6_3 = m_{6cm}/m_{3cm}$.
We also used the position angles at the two frequencies to derive the two-point
Faraday rotation ($RM$). The $RM$ is defined as $\chi(\lambda^2) =
\chi_{intr}+RM \lambda^2$ where $\chi_{intr}$ is the intrinsic position angle
and $\chi$ the apparent position angle at the $\lambda$ of the observations. 
Since the position angles are ambiguous by $n\pi$, an ambiguity affects also
the $RM$ calculated from only two frequencies. Thus, the $RM$ presented here
will have to be confirmed by observing the objects at at least one other
frequency.

\subsection{VLA Observations}

VLA snapshot observations were obtained for 8 sources in the declination range
north of $-30\degr$ using the A-array configuration on the 26 Oct 96.
The VLA observations are summarised in Table~3.  The observations were carried
out using the standard 6~cm (4.8~GHz) continuum mode, that is, with two
50~MHz-wide channels at bandwidth-separated frequencies (4.835 \& 4.885 GHz).
Each source was observed in two scans of about 15 minutes.

Data reduction followed standard procedures using the NRAO software package
AIPS.  The flux scale was calibrated against observations of the flux
standards 3C~48 \& 3C~286, adopting the standard scaling of Baars (1977).  A
few iterations of phase-only self-calibration were applied to each source. 
Again, together with the total intensity images we have also obtained the
images for the Stokes parameters, the polarised intensity ($P$) and
position-angle ($\chi$) images. 

The VLA observing parameters are summarised in Table~3.

\begin{table*}
\voffset=1truecm
\begin{center}
{\bf Table 2:  ATCA observations}

\medskip
\def\mc{\multicolumn{1}{c}}
\def\mcc{\multicolumn{2}{c}}
\def\mccc{\multicolumn{3}{c}}

\begin{tabular}{lccccccc} \hline\hline
 Object   & Conf. & $\lambda$  & \mccc{Resolution}         &  $\sigma_I$ & $\sigma_P$ \\
          &          &  cm     &  arcsec & arcsec  & deg   & mJy/beam    &      mJy/beam   \\
\hline
          &          &    &        &         &        &         &         \\
0039--44   & 6A       & 6  & 2.26   &  1.55   &  --6.2 & 0.26    &  0.19   \\
          &          & 3  & 1.29   &  0.91   &  --6.7 & 0.18   &  0.15 \\
0043--42  & 6A+1.5D  & 6  & 2.19   &  1.61   &  --4.8  & 0.48  &   ...      \\
          &          & 6* & 4.44   &  3.15   &  --3.0  & 0.60   & 0.15        \\
          &          & 3  & 1.19   &  0.88   &  --5.1 & 0.20   &  ...          \\
0409--75  & 6A       & 6  & 2.02   &  1.25   & --14.6 & 0.81   &   0.35         \\
          &          & 3  & 1.17   &  0.73   & --14.7 & 0.28   &   0.11         \\
0625--53  & 6A+1.5D  & 6  & 2.03   &  1.64   &   27.3 & 0.71   &   0.16       \\
          &          & 3  & 1.10   &  0.92   &  28.9  & 0.28   &   0.10        \\
1938--15  & 6A       & 6  & 5.89   &  1.47   & --2.4  & 0.60    &   0.12         \\
          &          & 3  & 3.41   &  0.88   & --1.8  & 0.40    &   0.11        \\
1954--55  & 6A       & 6  & 2.43   &  1.29   &  78.1  & 0.53   &   0.15         \\
          &          & 3  & 1.63   &  0.76   &  78.6  & 0.42   &   0.10       \\
2058--28  & 6A+1.5D  & 6  & 3.17   &  1.67   & --10.7 & 0.29   &   0.15       \\
          &          & 6* & 7.13   &  3.56   & --9.3  & 0.47   &   0.16       \\
          &       &       &       &        &       &       \\
\hline\hline
\end{tabular}
\end{center}

$^*$ Values derived  from the lower resolution map (3~km maximum baseline)

\end{table*}

\begin{table*}
\voffset=1truecm
\begin{center}
{\bf Table 3: VLA observations}

\medskip
\def\mc{\multicolumn{1}{c}}
\def\mcc{\multicolumn{2}{c}}
\def\mccc{\multicolumn{3}{c}}

\begin{tabular}{lccccccc} \hline\hline
 Object   & Conf. & $\lambda$  & \mccc{Resolution} &  $\sigma_I$ & $\sigma_P$ \\
          &     & cm     &  arcsec & arcsec  & deg   & mJy beam$^{-1}$  & mJy beam$^{-1}$  \\
\hline
          &     &      &        &         &        &         &       \\
0034--01  & A   & 6  & 0.41   & 0.38    & --33   &  0.04   & 0.03    \\
0035--02  & A   & 6  & 0.42   & 0.37    & --18   &  0.19   & 0.075   \\
0117--15  & A   & 6  & 0.46   & 0.31    &   1    &  0.11   & 0.06    \\
0442--28  & A   & 6  &  1.14  & 0.44    & --22   &  0.09   &0.032    \\
0453--20  & A   & 6  &  0.89  & 0.43    & --22   &  0.30   & 0.08    \\
1602+01   & A   & 6  &  0.39  & 0.36    & --38   &  0.12   &  0.085  \\
1938--15  & A   & 6  &  0.53  & 0.29    & --20   &  0.65   &  0.21  \\
2314+03   & A   & 6  &  0.40  & 0.36    & --42   &  0.09   & 0.029    \\
          &     &    &        &         &        &        &          \\
\hline\hline
\end{tabular}
\end{center}
\end{table*}

\section{Results}

Figs.  1 to 14 show the final images obtained for the galaxies observed with
ATCA. The electric field vectors are also shown (with the length proportional
to the fractional polarization). 
For 1938--15 we show the VLA image (see Fig.~9) together with the ATCA
images of this source.  The derived values of the fluxes and polarisation are
given in Table~4.  For the 3 cm they are obtained from the lower resolution
images.  Table~4 presents for the whole
source and other sub-regions (indicated in column 2): the fluxes at 6 and 3~cm
(columns 3 and 4 respectively); the fractional polarisation at 6 and 3~cm; the
depolarization, spectral index and $RM$ between these two frequencies.

Figs. 15 to 26 show the final images obtained for the galaxies observed with the
VLA. The parameters derived from these images are listed in Table~5. This
table presents the radio fluxes and the fractional polarisation at 6~cm for
the whole sources and their sub-regions indicated in column 2.
 
In the following section we describe in more details the results for the
single objects.

\subsection{Radio Galaxies Observed with ATCA}

{\bf 0039--44:} this is a relatively small radio source with
the emission dominated by the two lobes.  No core or jets have been detected. 
Both lobes are polarised with the eastern one showing an higher fractional
polarisation compared to the western side.  The eastern lobe is slightly
depolarised while the western side is consistent with no depolarization. 

In the optical, this galaxy has a high ionization emission line spectrum with
a strong blue, polarized continuum.  No extended emission lines have been
observed in this object. 

{\bf 0043--42:} this is a very extended source with a typical FRII morphology. 
It was previously mapped by Duncan \& Sproats (1992) and, at lower resolution
(843~MHz with the Molonglo Synthesis Telescope, MOST), by Jones \& McAdam
(1992). 

Unfortunately, at 3-cm most of the extended emission is resolved out at the
resolution of our data.  Thus, only the hot-spots could be imaged at this
frequency and they are shown in the inserts of Fig.  3.  At 6~cm no radio core
has been detected.  The polarisation is very uniform in the lobes with the
electric field well aligned with the position of the radio axis. 

In the optical it shows only weak, low-ionization emission lines and its
continuum appears to be typical of early-type galaxies.  Thus, this is an
example of FRII (and powerful) radio galaxies with significantly weaker
emission lines that expected from the radio power-emission line luminosity
correlation.

{\bf 0409--75:} this FRII is the highest red-shift object in the sample
and one of the most powerful source in the southern hemisphere (Alvarez et
al. 1993).  It was previously mapped by Duncan \& Sproats (1992). 

It is quite a small radio source  with the radio emission
dominated by two bright lobes.  Both lobes have an high depolarization,
slightly higher in the eastern lobe.  Also the
rotation measure is large especially in the eastern lobe. 

Interestingly, the ionization state is  low for such a powerful radio
galaxy: \Oiii\ is barely detected while \Oii\ is strong.  However
the continuum is bluer than an elliptical galaxy at the same red-shift (Dickson
1997).  According to Tadhunter et al. (1993) the \Oii\ line is extended
but this does not seem to be the case in \Oiii.

{\bf 0625--53:} This is a FRI galaxy associated with a dumbbell galaxy (Lilly
\& Prestage 1987, Gregorini et al.\ 1994) that is also the brightest member of
the cluster Abell 3391.  At the resolution of our observations, it shows a
wide-angle tail (WAT) structure with two tails sharply bending at about 20
arcsec north and 40 arcsec south of the nucleus.  This morphology is also
confirmed by the 13-cm ATCA image presented by Otani et al.\ (1998).  From an
X-ray/radio comparison, these authors found evidence for a possibly strong
interaction between the radio jets and the surrounding material.  The jets are
deflected possibly as the result of pressure gradients or winds in the
intracluster medium and they seem to `escape' into regions of lower X-ray
brightness.  This galaxy has also been observed at lower resolution by
Gregorini et al.\ (1994).  In their image, the northern tail is even more
prominent and a low-brightness diffuse region is also observed as a western
extension (for $\sim 45$ arcsec) to the southern lobe. 

The two lobes show a similar fractional polarisation at both 6 and 3 cm and
they both show no significant depolarisation.  The $RM$ is quite uniform in
the southern lobe but a large range of $RM$ values is observed in the northern
lobe (see Fig.8) with the central region around $RM \sim 350$ rad m$^{-2}$,
while the northern tail is showing values of $\sim -430$ rad m$^{-2}$.  There
is the possibility that the values are continuous at the step and the reason
for the `apparent' jump is the $n \pi$ ambiguity in the position angle: only a
study of the polarisation at other frequencies will be able to confirm this
jump in $RM$.  However, it may be worth noticing that, by looking at Fig.~7 in
Otani et al.  (1998) (i.e.  the radio/X-ray overlay), the jump in $RM$ occurs
in the region where the radio emission seems to `escape' the brighter X-ray
emission. 

{\bf 1938--15:} this is another small source dominated by the radio emission
from two lobes.  No core or jets have been detected.  It shows quite a strong
asymmetry in the depolarization: the eastern lobe has the lower fractional
polarisation and the stronger depolarization. 

This galaxy has a high ionization emission line spectrum.  From the new
optical spectra (Dickson 1997) this galaxy is now classified as a Broad Line
Radio Galaxy (BLRG) because it shows a prominent broad component in the
MgII$\lambda$2800\AA\ and H$\beta$.  As with many other BLRG, it is detected
in X-ray by ROSAT (Siebert et al.  1996). 

{\bf 1954--55:} this is an FRI source comprising two jet-like structures. 
When observed at low frequency and low resolution (Jones \& Mc Adams 1992),
this object is embedded in an elongated and bent low-brightness halo extended
more than 5 arcmin.  The integrated spectral index we find is much steeper
than from the single-dish observations (see Table~1) and this is probably due
to resolved emission in the 3~cm image. 

Neither jet shows evidence for significant depolarization.  The rotation
measure is quite uniform in the southern lobe ($\sim 126$ rad m$^{-2}$).  In
the northern lobe, most of the lobe shows a value of around 18 rad m$^{-2}$
except in a region (south-west) where the values are around $\sim -83$ rad
m$^{-2}$.

{\bf 2058--28:} the morphology of this FRI source is better defined than in the
previously available radio image. A strong jet is also now observed. The
difference with the previous radio image is likely due to the short
integration of the previous observations.  Unfortunately, this galaxy
could not be imaged properly at 3-cm and therefore we do not show the 3~cm
data here.  A low resolution radio image was obtained by 
Christiansen et al. (1977)
and shows that the source is embedded in a low brightness halo, extended more
than 8 arcmin, which is not visible in our image.

Two diffuse lobes without hot-spots are observed. The polarisation in
these lobes has a ring-like shape and the position angle of the electric
field is radial in both the lobes. 

2058--28 has a core flux density higher than in the previous VLA 6 cm data. 
The detected difference is likely to be due to the old value of the radio core
being unreliable, for the reasons mentioned above, more than variability in
the core flux.  Furthermore, the new core flux density is consistent with the
PTI data derived at 13 cm (Morganti et al.  1997a). 

In the optical it shows only weak, low-ionization emission lines and its
continuum appears to be typical of early-type galaxies.

\begin{table*}
\voffset=1truecm
\begin{center}
{\bf Table 4}

\medskip
\def\mc{\multicolumn{1}{c}}
\def\mcc{\multicolumn{2}{c}}
\def\mccc{\multicolumn{3}{c}}

\begin{tabular}{lcccccccc} \hline\hline
 Object   &      & $S_{6cm}$  & $S_{3cm}$ &  $m_{6cm}$ & $m_{3cm}$  &
DP$^{6cm}_{3cm}$ & $\alpha^{6cm}_{3cm}$ & RM  \\
         &         & Jy    &   Jy    &  \%   &  \%  &   &   & rad m$^{-2}$      \\
\hline
0039--44 & Total   & 1.169 & 0.606   & 7.5   &  6.7 & ...  &  1.20  &  ...   \\
         & E lobe  & 0.678 & 0.344   & 8.0   & 10.2 & $0.79$ & 1.24 & $-27$  \\
         & W lobe  & 0.486 & 0.261   & 3.7   &  2.7 & $1.35$ & 1.14 & $-32$  \\
         &         &       &       &       &      &      &      &       \\
0043--42$^*$ & Total   & 2.776 & ...  & 30.3  & ... & ...  & ...  &  ...  \\
         & HS North    & 1.182 & ...  & 20.1  & ... & ...  & ...  &  ...  \\
         & Ext. North  & 0.507 & ...  & 37.1  & ... & ...  & ...  &  ...  \\
         & HS South    & 0.397 & ...  & 20.0  & ... & ...  & ...  &  ...  \\
         & Ext. South  & 0.550 &  ... & 28.9  & ... & ...  & ...  &  ...  \\
         &             &       &      &       &     &      &      &       \\
0409--75 & Total   & 4.508  & 2.315  & 2.5   & 6.6  & ...  & 1.22 &  ...   \\
         & E lobe  & 2.671  & 1.459  & 1.1   & 3.4  & 0.30 & 1.10 & +522     \\
         & W lobe  & 1.843  & 0.895  & 4.5   & 9.3  & 0.46 & 1.32 & +235     \\
         &         &       &       &       &      &      &      &       \\
0625--53 & Totale  & 1.543 & 0.787 &15.7   & 15.1 & 0.96 &  1.23  &  +150     \\
         & Core    & 0.025 & 0.019 & ...   & ...  & ...  &  0.47  &  ...     \\
         & N Lobe  & 0.747 & 0.383 & 12.1  & 14.8  & 0.91 & 1.22 &  +43$^a$  \\
         & S Lobe  & 0.691 & 0.351 & 12.6  & 12.3 & 1.02 & 1.24   & +293    \\
         &         &       &       &       &      &      &      &       \\
1938--15 & Total   & 2.377  & 1.263  & 6.3  & 8.8  & ...  & 1.16  & ...     \\
         & E lobe  & 1.370  & 0.721  & 2.1  & 3.8  & 0.53 & 1.17 & --105  \\
         & W lobe  & 1.032 & 0.549  & 11.6  & 13.7 & 0.98 & 1.15 & --79  \\
         &         &       &        &       &      &      &      &       \\
1954--55 & Total   & 1.630  & 0.559  & 32.7  & 34.0  & ... & 1.95  & +33     \\
         & Core    & 0.055  & 0.052  & ...   &  ...  & ... & 0.10  & ...   \\
         & N lobe  & 1.340  & 0.351  & 39.1  & 39.8 & 0.98 & 2.45  &  0$^a$ \\
         & S lobe  & 0.608 & 0.165  & 26.5  & 23.2 & 1.06 & 2.38   & 126  \\
         &         &       &       &       &      &      &      &       \\
2058--28$^*$ & Total & 1.216 & ...  & 28.0  & ...  & ...  & ...  & ...  \\     
            & Core  &  0.123 & ...  & ...   & ...  & ...  & ...  & ...  \\
          & S lobe  &  0.591 & ...  & 25.3  & ...  & ...  & ...  & ...  \\
          & S jet   &  0.069 & ...  & 13.4  & ...  & ...  & ...  & ...  \\
          & N lobe  &  0.440 & ...  & 33.4  & ...  & ...  & ...  & ...  \\
          &         &        &      &       &      &      &      &      \\
\hline\hline
\end{tabular}
\end{center}

$^*$ Values derived  from the lower resolution map (3~km maximum baseline). \\
$^a$ See notes to the sources (\S 3.1)
\end{table*}

\subsection{Radio Galaxies Observed with VLA}

{\bf 0034--01 (3C15):} this source is now resolved into a prominent one-sided
jet, while only some structure is detected in the two lobes (due to the
relatively high resolution).  The southern lobe is the brighter one but only
shows a `warmer' spot (as perhaps expected looking at the low resolution image
in Morganti et al.  1993).  This galaxy has been observed at 3.6 cm by Leahy
et al.  (1997) who define its structure as intermediate between FR classes I
and II of FR, although its radio luminosity is typical of class II galaxies
(indeed it was classified as FRII in Morganti et al.  1993).  Knots are
detected along the jet as in the 3.6~cm images.  

The core flux density measured in the new image is much lower than from the 6
cm VLA data: this is probably due to the fact that low resolution of the
previous observations (about 3 arcsec) included the base of the strong jet in
what was defined as the core.  

The electric field in the jet is perpendicular to
the jet axis in the first blob (closer to the nucleus) and then become
parallel to the jet, although a different position angle seems to be characteristics of
the regions at the edge.  This is likely due to effect of shear layer as often
found in jets in FRI and described in detailed by Laing et al.  (1996). 

In the optical only a weak \Oiii\ emission has been detected from
this source (Tadhunter et al.
1993), and the continuum is typical of early-type galaxies.

{\bf 0035--02 (3C17):} this radio galaxy presented a very peculiar radio
morphology in our previous low resolution radio image (Morganti et al.  1993). 
The new high resolution image clarifys the real structure of this source.  On
the south-east side of the nucleus a very bent jet is observed.  The first
part of the jet is dominated by a bright blob.  On the western side a lobe
structure is observed with a ring-like shape, i.e.  because of the minimum in
intensity in the centre.  Also this could be a very bent jet seen in a
particular position angle.  A VLBA map (Venturi \etal\ 1996) shows as on the
milliarcsec scale a one-sided jet is observed in the eastern side and in the
same position angle  as the bright blob observed in our VLA map. 

In this galaxy, the H$\alpha$ emission line has a strong broad component
(Dickson 1997) and the \Oii\ and \Oiii\ emission lines are extended.
There is also evidence for significant optical polarization
in the nucleus of the galaxy (Tadhunter et al. 1997). 
This object has been detected in
X-rays (Siebert et al.  1996), as have most of the BLRG in our sample. 

Given the optical polarization properties (Tadhunter et al.  1997) this object
may be a BL Lac-type in which the jet is pointing close to the line of sight
(thus accentuating any wiggles in the jet).  It also looks very similar to the
southern BLRG/quasar PKS~2300-18 which has been interpreted in terms of
precession.

{\bf 0117--15 (3C38):} is a double lobed radio galaxy with a typical FRII
structure.  No core has been detected.  Both lobes have clear hot spots and
the southern lobe shows a clear bend in a direction almost perpendicular to
the line connecting the two lobes.  Similar structure detected, e.g., in 3C326
(Ekers et al.  1978) has been explained as result of a precessing beam.  An
alternative explanation has been suggested from a systematic study of bridges
in double radio sources by Leahy \& Williams (1984).  The sharp distortion
often observed in these sources can be expected as consequence of the
interaction between the strong back-flow and the galactic atmosphere if the
jet is much lighter than the confining medium.  

In the optical, this galaxy has a high ionization spectrum with both the
continuum and the line emission extended in the north-east direction.  The
continuum shows a strong UV/blue excess (Dickson 1997) and substantial optical
polarisation (Tadhunter et al.  1997).  The radio lobe on the same side as the
extended emission lines is closer to the nucleus and has the lower fractional
polarisation.  This may indicate a higher depolarization but confirmation will
require observations at different frequencies. 

{\bf 0442--28:} the new radio observations for this FRII radio galaxy shows a
complex morphology for its radio lobes.  A core has been detected but no radio
jets.  

This object has strong, narrow emission lines (Tadhunter et al.  1993).  It is
also detected in X-rays (Siebert et al.  1996), where it appears to be
extended, even though this source is not known to be associated with a
cluster. 

{\bf 0453--20:} from the new radio image this source appears as double lobed
with a clear jet in the north-west direction ending with an hot spot. In the
southern part no jet has been detected and a `warm-spot' is visible at the
edge of the lobe. The electric field is parallel to the direction of the jet
and becomes radial at the edge of the lobe.  

From Tadhunter et al. (1993) no optical lines are detected in this  galaxy 
and the continuum appears typical of early-type galaxies.

{\bf 1602+01 (3C327.1)} this source is dominated by a one sided knotty jet on
the south-east side.  This galaxy was also observed by Baum et al.  (1988)
with the B array at 6~cm, and by Hes (1995) at 8.4 GHz. 

In the radio, 1602+01 shows clear morphological similarities with 0035--02.
Similarities are also present in the optical spectrum of these objects.  In
fact, in the optical 1602+01 has a high ionization emission line spectrum
showing (as 0035--02) a broad component of the H$\beta$ emission (Dickson
1997).  The \Oii\ and \Oiii\ may be slightly extended.  As with most of the BLRG,
this galaxy has been detected in X-ray emission.

{\bf 1938--15:} this object has been also observed with the ATCA and described
above.  From the higher resolution of the VLA map (Fig.9) it is possible to
see that the western lobe is actually extended while the eastern one is only
slightly resolved.  Also at the higher resolution the east lobe shows a low
polarisation (as in the lower resolution ATCA image).  The west lobe (now
better resolved) shows  a higher fractional polarisation in the VLA image
($\sim 29$\%), indicating possible depolarization due to the beam at the
resolution of ATCA.

{\bf 2314+03 (3C459):} this is another small source of this sample also
studied before by Ulvestad (1985).  It is dominated by a strong core and two
lobes.  The eastern lobe is quite compact and much closer to the nucleus
compared to the western one.  A strong asymmetry in the polarisation can be
seen between the two lobes with the eastern one much less polarised (only 2\%)
than the western one consistent with what was found by Ulvestad (1985).  Also
in this object, as in 0034--01, the new core flux density is lower than in the
previous VLA data (Morganti et al.  1993).  Again, the difference is likely
due to the difference in resolution. 
  
In the optical this galaxy has a moderate ionization emission line spectrum and the
continuum is dominated by young stars.

\begin{table}
\voffset=1truecm
\begin{center}
{\bf Table 5}

\medskip
\def\mc{\multicolumn{1}{c}}
\def\mcc{\multicolumn{2}{c}}
\def\mccc{\multicolumn{3}{c}}

\begin{tabular}{lcccc} \hline\hline
 Object   &        &  S$_{6cm}$   &  $m_{6cm}$    & \\ 
          &        &    Jy    &   \%     & \\
\hline
0034--01  & core    &  0.032    &   ...     & \\
          & jet     &  0.292    &  25.0    & \\
          & N lobe  &  0.049    &   ...    & \\
          & S lobe  &  0.109    &  53.2    & \\
          &         &           &          & \\
0035--02  & Total   &  1.279    &  33.4    & \\
          & core    &  0.530    &  ...     & \\
          & E jet  &   0.475    &  30.0    & \\
          & W lobe  &  0.275    &  40.3    & \\
          &         &           &          & \\
0117--15  & Total   &  1.336    &   25.2   & \\
          & N-E lobe & 0.985    &  15.4    & \\
          & S-W lobe & 0.347    &  35.2    & \\
          &         &           &          & \\
0442--28  & core    &  0.039    &  ...     & \\
          & N lobe  &  0.573    &  25.6    & \\
          & S lobe  &  0.386    &  37.8    & \\
          &         &           &          & \\
0453--20  & Total   &  0.952    &  47.4    & \\
          & core    &  0.034    &   ...    & \\
          & N lobe  &  0.546    &  47.1    & \\
          & S lobe  &  0.323    &  48.4    & \\
          &         &           &          & \\
1602+01   & Total   &  0.945    &  ...     & \\
          & core    &  0.080    &  ...     & \\
          & E lobe  &  0.423    &  21.6    & \\
          & W lobe  &  0.441    &  16.7    & \\
          &         &           &          & \\
1938--15  & Total   &  2.947    &  11.4    & \\
          & E lobe  &  1.729    &   2.8   & \\
          & W hot-spot & 1.094  &  9.2    & \\
          & W lobe  &  1.269    &  28.5   & \\
          &         &           &          & \\
2314+03   & Total   & 1.294    &   14.4   & \\
          & core    & 0.425    &   2.0    & \\
          & E lobe  & 0.608    &   2.0    & \\
          & W lobe  & 0.260    &  19.2    & \\
\hline\hline
\end{tabular}
\end{center}
\end{table}

\section{Discussion \& Conclusions}

We have presented in this paper new, higher resolution radio images of a group
of 14 galaxies belonging to the 2-Jy sample of radio sources.  The new images
improve on the data already available for these objects and, in general, the
database that we are building up on the sample.  

Although the radio data presented here are mainly useful in the
context of the work we are doing on the sample, it is
still interesting to highlight some general results that can be derived from
the new data. These will be briefly described in the following sections.

\subsection{ The radio cores}

From our previous studies we have found that the core dominance [i.e.  ratio
between the core and the extended radio fluxes
$R=S_{core}/(S_{tot}-S_{core})$, in tests of unified schemes commonly used as
an indicator for the orientation of a source] appears to depend on both the
morphological classification (Morganti et al.  1995) and the optical
characteristics of the radio galaxies.  Among the FRII radio galaxies, most
Narrow Line Radio Galaxies (NLRGs) show low values of $\log R$, while BLRGs
tend to have large $\log R$ (Laing et al.  1994, Morganti et al.  1997a,
Hardcastle et al.  1998) supporting the idea that BLRGs are more beamed toward
us.

What do the new radio data tell us? In the observed sample we have three BLRG:
0035--02, 1602+01, 1938--15 (two of them only recently classified as BLRG). 
Both 0035--02 and 1602+01 have very prominent cores and the core flux
densities measured in the new observations are consistent with what was found
before.  In particular we find $R$=0.33 and $R$=0.061 for 0035--02 and 1602+01
respectively (if using the total fluxes taken from single dish observations,
Peacock \& Wall 1985).  On the other hand, in 1938--15 we do not detect any
core in our higher resolution map.  This sets an upper limit to the value of
$R$ (core flux $\ltae$ 1.5 mJy; R$\sim$0.0004), well below the average value
for BLRG ($\sim 0.027$).  Thus, this object appears to be an exception like
0347+05 ($R_{2.3GHz} < 0.0006$) as pointed out by Tadhunter et al.  (1998). 
These objects deserve some follow-up work to understand their real nature and
why they show such differences from the other BLRG.

\subsection{Asymmetry in depolarization}

For the objects for which we have two frequencies available, i.e.  the objects
observed with ATCA, we could investigate the presence of asymmetries in the
depolarization.  This asymmetry has been found in 0039--44, 0409--75 and
1938--15 and it is also notable that these are among the
highest redshift sources in the sample and they are all small (i.e. $\ltae 70$
kpc) radio sources. 
No obvious correlation has been found between the depolarization and the $RM$
(but see later for 0409--75).  

If the depolarisation is produced by an external Faraday screen (as generally
believed), this screen can be due to either the X-ray halo around the radio
source or to gas associated with the radio galaxy itself.  
In the latter case, we do not have the resolution high
enough (specially in the ATCA data) to investigate in detail the structure of
DP and $RM$ in the sources.

On the other hand, if the depolarisation is due to the X-ray halo around the
radio source we can estimate the Faraday dispersion ($\Delta$) for our
small sources from the formula given in Garrington et al. (1991):
$DP = exp[-2k^2\Delta^2(\lambda_1^2-\lambda_2^4)/(1+z)^4]$ where DP is the
depolarization and $k=0.81$ if we want to obtain $\Delta$ in units of cm$^{-3}$ $\mu$G pc.
For our frequencies we derive $\Delta = 270 (1+z)^2(-\ln DP)^{1/2}$ and we
find values of the Faraday dispersion ranging from  $\sim 800$ cm$^{-3}$ $\mu$G pc
for the E lobe of 0409--75 to $\sim 240$ cm$^{-3}$ $\mu$G pc for the E
lobe of 0039--44. These values are typically higher than what found for
extended sources (Garrington et al. 1991, Morganti et al. 1997b) but they lie
on the established trend found by Garrington \& Conway (1991) between the
linear size and the Faraday dispersion as do the CSS (Garrington \& Akujor
1996).  For the CSS, this result is expected if they represent simply young
version of the extended radio galaxies but living in a similar kind of
environment. In this case, the trend between size and $\Delta$ would represent
the radial decline of the density of the environment and the small
radio sources in our sample could be, to first order, seen as an evolved phase
of CSS on their way to becoming extended objects.

Of the three objects considered above, the only case in which both a large
depolarization and a large $RM$ has been observed is 0409--75.  It is
interesting to note that this object also show a quite low ionization state
(e.g.  \Oiii/\Oii\ = 0.19, Dickson 1997) for a galaxy of such a high radio
power.  Thus, in the case of this object the weakness of \Oiii\ could be due
to a low ionization parameter, for instance, due to an interaction between the
radio jet and a particularly rich ambient gas.


\begin{thebibliography}{}
\bibitem{} Alvarez H., Aparici J., May J., Navarrete M. 1993, A\&A 271, 435
\bibitem{} Baars et al. 1977, A\&A 61, 99
\bibitem{} Baum S.A., Heckman T. Bridle A.H., van Breugel W., Miley G. 1988,
ApJS, 68, 643 
\bibitem{} Christiansen W.N., Frater R.H., Watkinson A., O'Sullivan J.D.,
Lockhart I.A., Goss, W.M. 1977, MNRAS, 181, 183
\bibitem{} Dickson 1997, PhD Thesis, University of Sheffield
\bibitem{} Duncan \& Sproats 1992, PASA 10, 16
\bibitem{} Ekers R.D., Fanti R., Lari C., Parma P. 1978, Nature 276, 588
\bibitem{} Fanaroff B.L., Riley J.M. 1974, MNRAS 167, 31p
\bibitem{} Fanti C., Fanti R., Dallacasa D., Schilizzi R.T., Spencer R.E. \&
Stanghellini C. 1995 A\&A 302, 317;  
\bibitem{} Laing R.A., Jenkins C.R., Wall J.V., Unger S.W. 1994, in ``The
Physics of Active Galaxies'', Bicknell G.V., Dopita M.A., Quinn P.J., eds., 
ASP Conf.Ser. 54, p227;  
\bibitem{} Laing R.A. 1996, in ``Energy Transport in Radio Galaxies and
Quasars'', Hardee P.E., Bridle A.H., Zensus J.A., eds., ASP Conf. Series, 
Vol. 100, p.241
\bibitem{} Leahy J.P., Williams A.G. 1984, MNRAS 210, 929
\bibitem{} Leahy J.P. et al. 1997 MNRAS 291, 20
\bibitem{} Lilly S.J. \& Prestage R.M. 1987, MNRAS 225, 531 
\bibitem{} Garrington S.T., Conway R.G. \& Leahy J.P. 1991, MNRAS 250, 171
\bibitem{} Garrington S.T., Conway R.G. 1991, MNRAS 250, 198
\bibitem{} Garrington S.T., Akujor C.E. 1996, in ``Extragalactic radio
sources'', Ekers R., Fanti C., Padrielli L. (eds.) Kluwer Academic Publishers,
p. 77
\bibitem{} Gregorini L., de Ruiter H.R., Parma P., Sadler E.M., Vettolani G. \&
Ekers R.D. 1994, A\&AS 106, 1
\bibitem{} Hardcastle M.J., Alexander P., Pooley G.G., Riley J.M., 1998, MNRAS
296, 445
\bibitem{} Hes R. 1995, PhD Thesis, University of Groningen
\bibitem{} Jones P.A. \& McAdam W.B. 1992, ApJS, 80, 137
\bibitem{} Morganti R., Killeen N.E.B. \& Tadhunter C.N. 1993, MNRAS, 263, 1023
\bibitem{} Morganti R., Oosterloo T.A., Fosbury R.A.E. \& Tadhunter C.N. 1995
MNRAS 274, 393  
\bibitem{} Morganti R., Oosterloo T.A., Reynolds J., Tadhunter C.N.  \&
Migenes V. 1997a, MNRAS 284, 541 
\bibitem{} Morganti R.  \& Tadhunter C.N.  1997, Mem.  S.A.It.  Vol. 68, 243
\bibitem{} Morganti R., Parma P., Capetti A., Fanti R., de Ruiter H.R.,
Prandoni I. 1997b, A\&AS 126, 335
\bibitem{} Padovani P., Morganti R., Siebert J., Cimatti A., Vagnetti F.  1999,
MNRAS 304, 829
\bibitem{} Prestage R.M., Peacock J.A. 1983, MNRAS 204, 355
\bibitem{} Reynolds J.E. 1996 in {\it ``Australia Telescope Compact Array, Users's
Guide''} eds. W.M. Walsh \& D.J.McKay
\bibitem{} Sault, R.J., Teuben, P.J., Wright, M.C.H.  1995 in {\it ``Astronomical
Data Analysis Software and Systems IV''}, eds.  R.  Shaw, H.E.  Payne
and J.J.E.  Hayes, Astronomical Society of the Pacific Conference
Series, 77, 433
\bibitem{} di Serego Alighieri S., Danziger J., Morganti R., Tadhunter C.N.
1994, MNRAS 269, 998
\bibitem{} Siebert J., Brinkmann W., Morganti R., Tadhunter C., Danziger J., Fosbury R. \&
di Serego Alighieri 1996, MNRAS 279, 1331
\bibitem{} Tadhunter, C.N., Morganti, R., di Serego Alighieri, S., Fosbury, 
R.A.E. \& Danziger, I.J.  1993, MNRAS, 263, 999
\bibitem{} Tadhunter C.N., Dickson R., Morganti R., Villar-Martin M.  1997,
in proceeding of ESO/IAC workshop ``Quasar Hosts''
Clements \& Perez-Fournon (eds.)  p.311
\bibitem{} Tadhunter C.N., {  Morganti R.}, Robinson A., Dickson R.,
Villar-Martin M. \& Fosbury R.A.E. 1998, MNRAS 298, 1035
\bibitem{} Trussoni E., Vagnetti F., Massaglia S., Feretti L., Parma P., Morganti
R., Fanti R., Padovani P.  1999, A\&A 348, 437
\bibitem{} Ulvestad J.S. 1985, ApJ 288, 514
\bibitem{} Venturi T., Morganti R., Tzioumis, A et al. 1996, 3rd Asia-Pacific
Telescope Workshop, ed. E.King,  p.287    
\bibitem{} Wall, J.V. \& Peacock, J.A. 1985, MNRAS, 216, 173.

             
\end{thebibliography}
\end{document}